# Empty space-time, general relativity principle and covariant ether theories

Alexander L. Kholmetskii
Department of Physics, Belarus State University,
4, F. Skorina Avenue, 220080 Minsk Belarus
E-mail: kholm@bsu.by

We look for the properties of empty space-time proceeding from the general relativity principle. An infinite number of the so-called covariant ether theories (CETs) has been found, which, like the special relativity theory (SRT), satisfy all known experimental facts in the physics of empty space-time. In this connection a new approach to the problem of experimental testing of SRT is discussed. In particular, we show that covariant ether theories predict a dependence of Thomas-Wigner angle on an "absolute" velocity of a reference frame of observation. Hence, a measurement of this dependence is capable to distinguish SRT and CETs. It has been shown that the Lorentz ether theory is one of CETs, corresponding to the admissible Galilean transformations in physical space-time. Hence, we conclude that SRT and Lorentz ether theory can be distinguished experimentally, at least in principle. A crucial experiments, based on the Mössbauer effect, has been proposed.

## 1. Introduction

Modern physics accepts two relativity principles: the special relativity principle (SRP) asserting that fundamental physical equations do not change (they are form-invariant) under transformations between inertial reference frames in an empty space, and the general relativity principle (GRP) stating that fundamental physical equations do not change their form (they are covariant) under transformations between any frames of references. Strictly speaking, the GRP requires a covariance of physical equations with respect to "admissible" space-time transformations, which keep the requirements $g_{00}>0$, $g_{ij}dx^i dx^j<0$, where $g$ is the metric tensor, and $i, j=1...3$.

    Both relativity principles were introduced by Einstein at the beginning of 20[th] century: SRP lied on the basis of special relativity, while GRP in combination with the equivalence principle gave rise to general relativity theory.

    The GRP is one of the deepest principles of physics and it means that any phenomenon can be described from any reference frame, which can be realized in nature [1-3]. A problem of experimental test of GRP has no further meaning than a problem of experimental testing of space-time homogeneity, causality principle, etc. The mentioned principles constitute the corner stones of modern knowledge, and we simply accept their validity: otherwise the whole of modern physics would be destroyed.

    A fundamental physical consequence of SRP is an impossibility to reveal an absolute velocity of an inertial reference frame. There is a widespread opinion that SRP is a direct consequence of GRP in the case of inertial motion in an empty space. If it were actually so there would be no meaning to testing SRP experimentally. Indeed, in such a case an experimental test of SRP would mean simultaneously a test of GRP, and such experiments seem to be impractical. However, the SRP in not, in general, a consequence of GRP; it represents an independent physical assumption. Only in the case when empty space-time has a pseudo-

Euclidean geometry with Minkowskian metric[1] in any inertial frame (which special relativity theory demands) can we derive a form-invariance of physical equations with respect to the Lorentz transforms as the special inference from the covariance principle. At the same time, from the viewpoint of formal logic we may introduce into consideration ether theories, where a metric in an arbitrary inertial frame depends on its velocity in the ether. If, nevertheless, the metric coefficients in such a co-ordinate geometry continue to be "admissible", such a theory would be in agreement with GRP, but in contradiction with SRP.

The present paper has a goal to inspect more closely the relationship between SRP and GRP, as well as to analyze the experimental facts from this point of view. Section 2 derives some important consequences of GRP when applied to empty space. Section 3 investigates the properties of hypothetical empty space-time with metrics that differ from Minkowskian, and Section 4 describes the covariant ether theories (CETs), which are developed on the basis of GRP and symmetries of an empty space-time. Section 5 presents possible experiments for verification of CETs, based on the Mössbauer effect. Finally, Section 6 contains some conclusions.

## 2. About some consequences of GRP for the case of inertial motion in an empty space

Let us write for an empty space a space-time transformation between two inertial reference frames in the general form

$$x_a = A_{ab} x'^b, \qquad (1)$$

where $a,b=0...3$, and $x$, $x'$ are four-vectors in the inertial frames. It is known that the principle of space-time homogeneity ensures linearity of this transformation [4]. The GRP requires that the transformations **A** constitute a group of Lee with ten parameters: four initial space-time coordinates, three Eulerian angles, and three projections of a relative velocity [4]. Further, let us exclude the trivial space translations and rotations. In such a case the transformation depends upon a single vectorial parameter - relative velocity $\vec{v}$, i.e.

$$x_a = A_{ab}(\vec{v}) dx'^b. \qquad (2)$$

The GRP also requires validity of the reciprocity principle [5]: the mutual velocities of two inertial reference frames should differ only by sign

$$\mathbf{A}^{-1}(\vec{v}) = \mathbf{A}(-\vec{v}). \qquad (3)$$

In its turn, the reciprocity principle ensures that $\det\mathbf{A}=1$ [5]. Thus, the transformations **A** are special orthogonal.

In fact, this is all that we can say about the properties of the matrix **A** proceeding from GRP and the principles of symmetry of empty space-time. In order to determine **A** in closed form, it is necessary to define a model of an inertial reference frame and to make some additional physical assumptions. (For example, under Einstein's postulates, the matrix **A** is equal to the Lorentz matrix **L** in Cartesian inertial reference frames). For these reasons nobody tried to pursue an analysis of the properties of empty space-time within GRP. However, as we will see below, poor information already obtained on the basis of GRP about the matrix **A** becomes nevertheless sufficient to determine a number of general laws of inertial motion. Such an analysis seems important for a better understanding of the experimental basis of SRT. Indeed, often an experiment dealing with inertial motion in empty space is unambiguously considered as a test of SRP. However, if we show that the result of this or that experiment can be explained by the GRP solely, that it means a test of GRP, not SRP, then such an experiment becomes useless for physics.

---

[1] By definition, the Minkowskian metric tensor has the form $g_{00}=1, g_{11}=g_{22}=g_{33}=-1$, and all others $g_{ab}=0$, where $a, b=0...3$.

In order to continue this consideration further, let us take a hypothetical assumption about existence of an "absolute space", which has pseudo-Euclidean geometry with a Minkowskian metric. We designate a preferred frame, attached to the "absolute space", as $K_0$. We look for a possibility to measure the absolute velocity $\vec{v}$ within a moving frame K for space-time transformation **A** in Eq. (2) with the properties defined above.

Consider two inertial frames $K_1$ and $K_2$ initially both resting in the absolute frame $K_0$. The frame $K_1$ contains some device D to measure the absolute velocity of that frame (by means of internal measuring procedures). Initially, $v=0$, and hence, the above mentioned device D stays in some state corresponding to $v=0$. Let us imagine that the frame $K_2$ acquires some constant absolute velocity $-\vec{v}$, see Fig. 1, a. Such an operation does not influence our device D in $K_1$, and thus, it remains in the original state. According to Eqs. (2), (3) one can denote a transformation from $K_1$ to $K_2$ as $\mathbf{A}^{-1}(-\vec{v})=\mathbf{A}(\vec{v})$ and to conclude that $\mathbf{A}(\vec{v})$ has no effect on the device D.

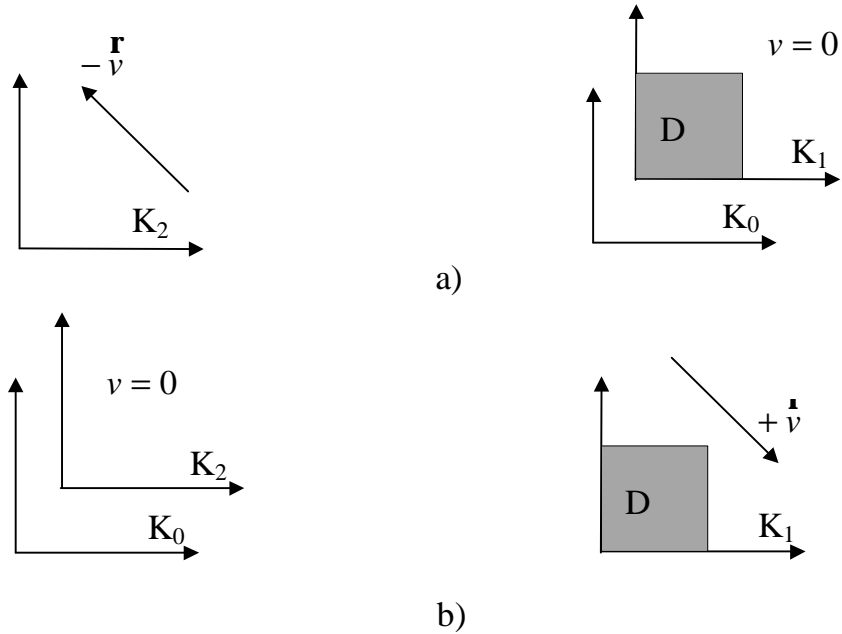

Fig. 1: a – the frame $K_1$ remains at rest in $K_0$, while the frame $K_2$ acquires a constant velocity $-\vec{v}$ in $K_0$; b – the frame $K_2$ rests in $K_0$, and $K_1$ moves at constant "absolute" velocity $+\vec{v}$ in $K_0$.

One can consider now a different case. Namely, the frame $K_2$ remains at rest in the absolute frame $K_0$, while the frame $K_1$ containing the device D acquires a constant velocity $+\vec{v}$ in the frame $K_0$ (see, Fig. 1, b). For such a case, a transformation from $K_1$ to $K_2$ (and $K_0$ as well) takes on the same form $\mathbf{A}(\vec{v})$, as for Fig. 1, a. On the other hand, it has been found that the $\mathbf{A}(\vec{v})$ has no effect on the state of the device D. Thus, according to the GRP, an absolute velocity is not observable in this kind of experiments. One can easily see that such a kind of experiments corresponds to a case where all inertial parts of the device D rest in the (laboratory) frame $K_1$.

Thus, we get the **first general inference of GRP** with respect to inertial motion in an empty space: **no absolute motion with a constant velocity could be detected by a device having all inertial parts resting with respect to one another.** This theorem explains the null results of all interference experiments searching for "ether wind", beginning with Michelson-Morley.

Let us consider now a device having inertial parts moving at constant non-zero relative velocities one versus another in the frame $K_1$. Each part having a different constant velocity $\vec{u}_i$ in $K_1$ could be attached to its own proper inertial frame $K_i$. Then for the first motion diagram ($K_1$ at rest in $K_0$, $K_2$ moving at the constant velocity $-\vec{v}$ in the absolute frame $K_0$, Fig. 2, a), a transformation from each $K_i$ to $K_2$ takes on the form $\mathbf{A}^{-1}(-\vec{v})\mathbf{A}(\vec{u}_i)=\mathbf{A}(\vec{v})\mathbf{A}(\vec{u}_i)$. Due to the fact, that a motion of the frame $K_2$ has no effect on the D (belonging to $K_1$) one can conclude that this transformation leaves all parts of the device D in the original state, regardless of the value of the index $i$, details of the transformation itself and particular construction of the device. In order to find the indication of D for the second motion diagram ($K_2$ at rest in $K_0$, $K_1$ moving at the constant velocity $+\vec{v}$ in $K_0$, Fig. 2, b), one can assume, in general, two kinds of transformations from each $K_i$ to $K_2$ (and $K_0$ as well): $\mathbf{A}(\vec{v})\mathbf{A}(\vec{u}_i)$ and $\mathbf{A}(\vec{v}\oplus\vec{u}_i)$. These transformations, generally, are not equal to each other, since orthogonal transformations are not commutative (the group of space-time transformations is non-Abelian). We already proved above that the first $\mathbf{A}(\vec{v})\mathbf{A}(\vec{u}_i)$ transformation does not change the state of D. Simultaneously we conclude that the other transformation $\mathbf{A}(\vec{v}\oplus\vec{u}_i)$, being different from $\mathbf{A}(\vec{v})\mathbf{A}(\vec{u}_i)$, changes the state of the measuring device D. Therefore, it is able to describe the difference of indications of D under "absolute rest" and "absolute motion" of the inertial reference frame $K_1$. This again shows that the GRP and SRP (where such a situation is impossible) represent two independent physical assumptions: generally speaking, the GRP does not forbid the existence of absolute space.

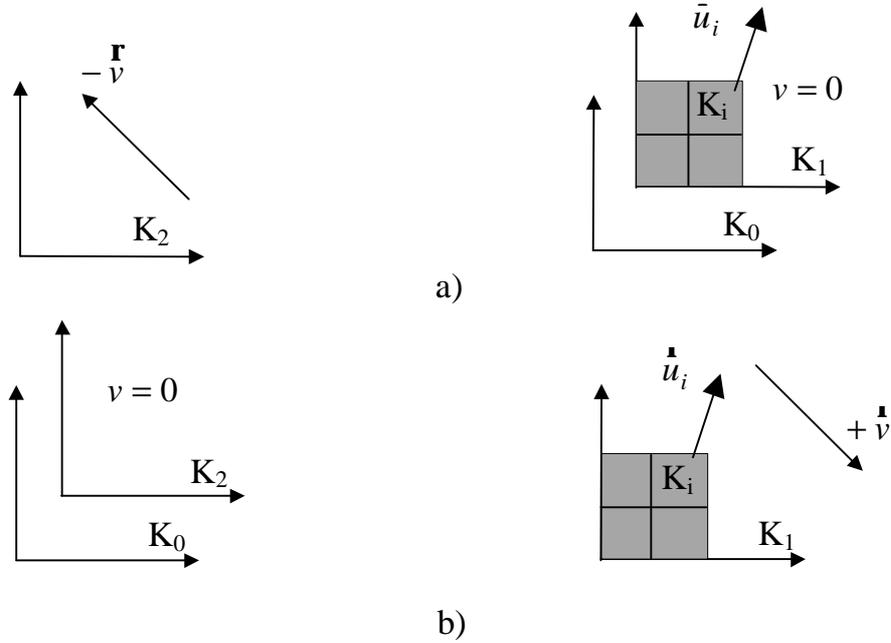

Fig. 2: a – the frame $K_1$ rests in $K_0$, and the frame $K_2$ acquires a constant "absolute" velocity $-\vec{v}$ in $K_0$; b – the frame $K_2$ remains at rest in $K_0$, while $K_1$ moves at a constant "absolute" velocity $+\vec{v}$ in $K_0$. A measuring instrument D contains the moving elements to be attached with the inertial reference frames $K_i$.

There is only one particular case ($\vec{v}$ is collinear to all $\vec{u}_i$) where $\mathbf{A}(\vec{v})\mathbf{A}(\vec{u}_i)=\mathbf{A}(\vec{v}\oplus\vec{u}_i)$, and the state of the device D does not depend on the absolute velocity $\vec{v}$ of the frame $K_1$ regardless of the particular construction of D.

Hence, we get the **second general implication of GRP** with respect to inertial motion in an empty space: **no absolute motion with a constant velocity could be detected by a device whose inertial parts move parallel (or opposite) to the absolute velocity.**

One can show that two general inferences of GRP obtained above, taken together, explain the null results of all experiments searching for "ether wind" velocity, performed up to date [6].

We stress that the first and second inferences of GRP were obtained in quite general form, and they do not depend on a specific construction of the device D, as well as a specific choice of the transformation **A**.

**Independence of the obtained inferences from any specific structure of the device D** has a principal importance. It means that a device D can be constructed on the basis of any known form of interaction, and it can be either macroscopic or microscopic. Hence, both inferences obtained above belong not only to relativistic kinematics, but remain valid for any other area of physics.

**Independence of the obtained inferences from a particular choice of the transformation A** in Eq. (2) seems to contradict the mainstream opinion that the Lorentz transformations **L** exclusively describe phenomena with non-observable absolute velocity in an empty space-time. However, our consideration of general motion diagrams in Figs. 1 and 2 indicates that such an opinion is erroneous. Further we will actually show that all available experimental facts in the physics of empty space can be explained under any admissible transformation **A**. A concurrent problem is to understand the physical meaning of ether space-time theories with different transformations **A**, as well as different successions of their actions for the motion diagram in Fig. 2, b. In order to solve these problems, it is necessary to analyze more closely the properties of empty space-time under an ether hypothesis.

## 3. Pseudo-Euclidean empty space-time with oblique-angled metrics

We will further analyze the ether theories, which adopt pseudo-Euclidean geometry with Minkowskian metrics for "absolute space". Since the motion of an arbitrary inertial frame does not influence geometry of absolute space, it continues to be pseudo-Euclidean for any moving inertial observer. However, due to possible dependence of space and time intervals on the absolute velocity, admitted in the ether theories, the metric coefficients also depend on the absolute velocity, and the metric tensor *g* in moving frames is no longer Minkowskian. This means for pseudo-Euclidean geometry that physical space-time four-vectors in an arbitrary inertial frame should be linear functions of Minkowskian four-vectors $x_L$:

$$(x_{ph})_a = B_{ab}(x_L)^b, \tag{4}$$

where the coefficients $B_{ab}$ do not depend on space-time coordinates of a moving inertial frame; they depend only on its absolute velocity $\vec{v}$. (This statement follows from the space-time homogeneity [6]). Such a kind of pseudo-Euclidean geometry has the so-called oblique-angled metric. Here $x_L$ obey the Lorentz transformation **L**:

$$x_{La} = L_{ab} x_L^{'b}. \tag{5}$$

In analysis of space-time with an oblique-angled metric, there is an essential methodological feature that has to be taken into account. Although this feature was stressed many years ago by Reichenbach (*e.g.*, [7]), present ether theories do not take it into account explicitly.

It may be natural to believe that in any inertial reference frame we are able to construct a method for measurement of space and time intervals such that the result of measurement directly gives the physical magnitude of the corresponding interval. But strictly speaking, this is a property exclusive to pseudo-Euclidean geometry with Minkowskian metrics. Only in this kind of geometry can we omit a distinction between physical space-time four-vectors and four-vectors, obtained via measurements [3, 6, 8, 9]. That is, in general, we have "measured" $x_{ex}$ and physical $x_{ph}$ four-vectors, and only in pseudo-Euclidean geometry with Minkowskian metrics do we have

$$x_{ph} = x_{ex} = x_L. \tag{6}$$

The essential property of space-time with oblique-angled metrics is the difference between measured and physical space-time four-vectors in arbitrary inertial reference frames: $x_{ph} \neq x_{ex}$. The necessity to distinguish them can be easily demonstrated with the Fitzgerald-Lorentz contraction hypothesis, which was first invoked to explain the null result of the Michelson-Morley experiment. According to this hypothesis, if a rod initially at rest in the absolute frame has the length $l$, then under motion at a constant absolute speed $v$ along its axis, the length of the rod becomes $l\sqrt{1-v^2/c^2}$. However, due to proportional contraction of the unit scale in an attached inertial reference frame, an experimenter in this frame measures the same length $l$ as in the case $v=0$: Fitzgerald-Lorentz contraction is not observable. Thus, we see that the length of the rod in physical space-time is $l_{ph}=l\sqrt{1-v^2/c^2}$, while the measured length is equal to $l_{ex}=l$, and $l_{ph} \neq l_{ex}$. One can easy demonstrate that the same conclusion is valid for time intervals in an empty space-time with the oblique-angled metrics: $t_{ph} \neq t_{ex}$. Thus, the four-vectors in physical space-time (hereinafter "physical" four-vectors) are not equal to the four-vectors, whose components were obtained via a measurement of corresponding space and time intervals (hereinafter "measured" four-vectors). Hence, in any alternative to SRT theory we have to derive separately the transformation rules for both kinds of four-vectors, when the condition (6) remains valid only for absolute space:

$$(x'_{ph})_a \neq (x'_{ex})_a \neq (x'_L)_a. \tag{7}$$

(Hereinafter the primed four-vectors belong to the absolute frame). This problem will be considered in the next Section.

**4. Covariant ether theories**

First of all, we notice that Eq. (4) under the condition (7) means that the matrix **B** becomes equal to the unit matrix when $v=0$, and

$$(x_L)_a = (x_{ph})_a \, (v=0). \tag{8}$$

This allows one to rewrite Eq. (4) in the form

$$(x_{ph})_a (\vec{v}) = B_{ab}(\vec{v})(x_{ph})^b (v=0), \tag{9}$$

which clearly indicates a physical meaning of the matrix **B**: it describes a dependence of physical space and time intervals in a moving inertial frame on its absolute velocity $\vec{v}$.

Further, let us write a relationship between time components of the four-vectors $x_{ph}$ and $x_L$, proceeding from Eq. (4):

$$(x_{ph})_0 = B_{00}(x_L)^0 + B_{0i}(x_L)^i \tag{10}$$

($i$=1..3). For two events at a fixed spatial point ($(x_{ph})^i = 0$)

$$(x_{ph})_0 (\vec{v}) = B_{00}(x_L)^0 = B_{00}(x_{ph})^0 (v=0). \tag{11}$$

Hence, the coefficient $B_{00}$ describes the change of clock rate at a fixed spatial point under its motion at the constant absolute velocity $\vec{v}$. Such a change takes place for both standard and physical time intervals. Therefore, the measured time interval at a fixed spatial point is

$$(x_{\text{ex}})_0 = (x_{\text{ph}})^0 / B_{00}. \tag{12}$$

For time intervals at two different spatial points, separated by the distance $(x_{\text{ph}})_i$, one should write

$$(x_{\text{ex}})_0 = [(x_{\text{ph}})_0 + \Delta(x_{\text{ph}})_0] / B_{00}. \tag{13}$$

where $\Delta(x_{\text{ph}})_0$ is the error of synchronization of clocks separated by the distance $(x_{\text{ph}})_i$ in oblique-angled space-time. (It appears due to possible anisotropy of light velocity at different directions under Einstein's synchronization of distant clocks). The value of $\Delta(x_{\text{ph}})_0$ can be found from the equality

$$(x_{\text{ph}})_{02} = (x_{\text{ph}})_{01}/2 \tag{14}$$

(Einstein's synchronization method), where $(x_{\text{ph}})_{01}$ stands for the time for light propagation from the first clock $Cl_1$ (at the origin of coordinates) to the second clock $Cl_2$ (at the point $(x_{\text{ph}})_i$) and back according to $Cl_1$, while $(x_{\text{ph}})_{02}$ is the reading of $Cl_2$ at the moment of arrival of the light pulse. For oblique-angled space-time the propagation time of light from $Cl_1$ to $Cl_2$ $(x_{\text{ph}})_{0+}$ is not equal, in general, to the propagation time in the reverse direction $(x_{\text{ph}})_{0-}$. Hence, an implementation of the equality (14) is possible only in the case where the readings of both clocks at the initial moment of time differ by the value $\Delta(x_{\text{ph}})_0$, and

$$(x_{\text{ph}})_{01} = [(x_{\text{ph}})_{0+} + (x_{\text{ph}})_{0-}], \quad (x_{\text{ph}})_{02} = [(x_{\text{ph}})_{0+} + \Delta(x_{\text{ph}})_0], \tag{15}$$

Hence, with account of Eq. (14), we obtain:

$$\Delta(x_{\text{ph}})_0 = \frac{1}{2}[(x_{\text{ph}})_{0-} - (x_{\text{ph}})_{0+}]. \tag{16}$$

Expressions for $(x_{\text{ph}})_{0+}$ and $(x_{\text{ph}})_{0-}$ can be found from Eq. (4):

$$(x_{\text{ph}})_{0+} = B_{00}(x_{\text{L}})^0 + B_{0i}(x_{\text{L}})^i, \quad (x_{\text{ph}})_{0-} = B_{00}(x_{\text{L}})^0 - B_{0i}(x_{\text{L}})^i. \tag{17}$$

Substituting Eq. (17) into Eq. (16), one gets:

$$\Delta(x_{\text{ph}})_0 = -B_{0i}(x_{\text{L}})^i. \tag{18}$$

Further substitution of Eqs. (18) and (10) into Eq. (13) gives:

$$(x_{\text{ex}})_0 = (x_{\text{L}})_0. \tag{19}$$

Thus, we have derived an important result: for any ether theory, adopting a Minkowskian metric of absolute space, the measured time intervals always obey the Lorentz transformations.

Looking at Eq. (19), we may ask the following question: does this equality continue to be valid for space intervals, too? In another words, would we get the equality

$$(x_{\text{ex}})_i = (x_{\text{L}})_i \tag{20}$$

for arbitrary admissible matrix **B** in Eq. (4)? In general, it is not. Let us show that Eq. (20) is realized only in the case where the coefficients $B_{i0}=0$. Indeed, write a relationship between space components of the four-vectors $x_{\text{ph}}$ and $x_{\text{L}}$, proceeding from Eq. (4):

$$(x_{\text{ph}})_i = B_{ia}(x_{\text{L}})^a = B_{i0}(x_{\text{L}})^0 + B_{ij}(x_{\text{L}})^j. \tag{21}$$

Introducing a unit scale $(x_{\text{phu}})_i$ in physical space-time, we can write the similar relation:

$$(x_{\text{phu}})_i = B_{i0}(x_{\text{L}})^0 + B_{ij}(x_{\text{Lu}})^j, \tag{22}$$

where $(x_{\text{Lu}})^i$ is the corresponding unit scale in Minkowskian space. Dividing (21) by (22), one obtains:

$$\frac{(x_{\text{ph}})_i}{(x_{\text{phu}})_i} = \frac{B_{i0}(x_{\text{L}})^0 + B_{ij}(x_{\text{L}})^j}{B_{i0}(x_{\text{L}})^0 + B_{ik}(x_{\text{Lu}})^k}.$$

Taking into account the obvious equality for Minkowskian space

$$(x_{\text{L}})^j / (x_{\text{L}})^i = (x_{\text{Lu}})^j / (x_{\text{Lu}})^i,$$

we get after manipulations:

$$\frac{x_{\text{ph}\,i}}{x_{\text{phu}\,i}} = \frac{B_{i0}x_{\text{L}}^0 + B_{ik}x_{\text{L}}^k}{B_{i0}x_{\text{L}}^0 + x_{\text{Lu}}^i \left[B_{ii} + \frac{1}{x_{\text{Lu}}^i}\sum_{j\neq i} B_{ij} x_{\text{Lu}}^j\right]} = \frac{B_{i0}x_{\text{L}}^0 + B_{ik}x_{\text{L}}^k}{B_{i0}x_{\text{L}}^0 + x_{\text{Lu}}^i \left[B_{ii} + \frac{1}{x_{\text{L}}^i}\sum_{j\neq i} B_{ij} x_{\text{L}}^j\right]} =$$

$$= \frac{B_{i0}(x_{\text{L}})^0 + B_{ik}x_{\text{L}}^k}{B_{i0}(x_{\text{L}})^0 + x_{\text{Lu}\,i}(B_{lm}x_{\text{L}}^m)/x_{\text{L}\,i}}. \qquad (23)$$

Obtained Eq. (23) proves our statement. Indeed, under $B_{i0}=0$ it transforms to

$$\frac{x_{\text{ph}\,i}}{x_{\text{phu}\,i}} = \frac{x_{\text{L}\,i}}{x_{\text{Lu}\,i}},$$

which is equivalent to Eq. (20): the measured scale in oblique-angled space-time coincides with its value in Minkowskian space-time. Then Eqs. (19), (20) are written simultaneously as

$$(x_{\text{ex}})^a = (x_{\text{L}})^a, \qquad (24)$$

which means that a distinction of oblique-angled metrics in moving inertial frames from Minkowskian metrics is not experimentally observable. In another words, an observer in any inertial frame moving in absolute space sees the world, almost as in SRT, for an infinite set of ether space-time theories with $B_{i0}=0$. (However, this does not yet mean that SRT and all ether theories cannot be distinguished experimentally. This problem will be analyzed below).

Now let us determine a physical meaning of the equality $B_{i0}=0$ in Eq. (4). For this purpose we combine Eqs. (1), (5), (7) and derive:

$$(x_{\text{ph}})_a = A_a^g(\vec{v}) L_{gb}^{-1}(\vec{v})(x_{\text{L}})^b. \qquad (25)$$

Comparing Eq. (25) with Eq. (4), we find

$$B_{ab}(\vec{v}) = A_a^g(\vec{v}) L_{gb}^{-1}(\vec{v}). \qquad (26)$$

Substituting into Eq. (26) the known form of the matrix **L** (see, *e.g.* [1]), denoting $g = 1/\sqrt{1 - v^2/c^2}$, and using Eq. (3), one gets:

$$B_{00} = gA_{00}, \qquad (27a)$$

$$B_{i0} = 0, \qquad (27b)$$

$$B_{0i} = A_{0i} + A_{00} \times [\frac{v_i}{c^2}g + (\frac{1}{A_{00}^2} - 1)\frac{v_i}{v^2}(g-1)], \qquad (27c)$$

$$B_{ij} = A_{ij} + A_{i0}\frac{v_j}{v^2}(1 - \frac{1}{g}). \qquad (27d)$$

We see that the coefficient $B_{i0}$ is equal to zero for any matrix **A**. One can show that an adoption of the reciprocity principle (3) is essential for vanishing of $B_{i0}$.

**Thus, we conclude that the equality $B_{i0}=0$ represents an equivalent form of the reciprocity principle, resulted from GRP.**

Therefore, for any admissible transformation **B**, satisfying the GRP, an experimenter will not detect a distinction of oblique-angled metrics of his coordinate geometry from Min-kowskian metrics. That is why all the experiments for verification of Lorentz transforms (beginning with the Michelson-Morley experiment and finishing with the modern experiments for search of ether velocity [10-12]) find an infinite number of alternative explanations of their results. This conclusion is in full accordance with the implications of GRP obtained in Section 2 for the case of inertial motion in empty space-time.

We name the acceptable theories "covariant ether theories" (CETs). The exclusive place of SRT among all such CETs is defined by the fact that it directly asserts the equality of measured and physical space-time four-vectors, *i.e.*, the equality of the matrices **A** and **L**, which means a Minkowskian metric of physical space-time in any inertial reference frame. In the alternative assumption **A**≠**L**, metric of physical space-time is, in general, oblique-angled, and we must distinguish space-time transformations for physical and measured four-vectors:

$$(x_{ph})_a = A_{ab}(x'_{ph})^b, \quad (x_{ex})_a = L_{ab}(x'_{ex})^b, \qquad (28)$$

where the primed four-vectors, as before, belong to the absolute frame $K_0$. This means that these transformations do not yet solve the main kinematical problem (determination of space-time transformations between two arbitrary inertial frames): they act only in the special case, where one of the frames is absolute. In order to find a transformation between two arbitrary inertial frames K and K", we should write

$$(x_{ex})_a = L_{ab}(\vec{v}_1)(x'_{ex})^b; \quad (x''_{ex})_a = L_{ab}(\vec{v}_2)(x'_{ex})^b, \qquad (29)$$

$$(x_{ph})_a = A_{ab}(\vec{v}_1)(x'_{ph})^b; \quad (x''_{ph})_a = A_{ab}(\vec{v}_2)(x'_{ph})^b, \qquad (30)$$

where $\vec{v}_1, \vec{v}_2$ are the absolute velocities of the frames K and K", respectively. Eliminating four-vector $x'^b_{ex}$ from Eqs. (29), and $x'^b_{ph}$ from Eq. (30), we obtain general transformations for measured and physical space-time four-vectors in two arbitrary inertial frames:

$$(x_{ex})_a = L_{ab}(\vec{v}_1)[L^{-1}(\vec{v}_2)]^{bg}(x''_{ex})_g, \qquad (31)$$

$$(x_{ph})_a = A_{ab}(\vec{v}_1)[A^{-1}(\vec{v}_2)]^{bg}(x''_{ph})_g, \qquad (32)$$

where the matrix **A** can be taken in arbitrary admissible form. Thus in contrast to SRT, under the hypothesis **A**≠**L**, Nature does not "know" a direct relative velocity of two arbitrary inertial frames K and K": it is always composed as a sum $\vec{v}_1 \oplus \vec{v}_2$, where $\vec{v}_1$ and $\vec{v}_2$ are the corresponding velocities of K and K" in the absolute frame $K_0$. This means, in particular, that direct rotation-free Lorentz transformation between measured space-time four-vectors in K and K" is impossible: according to general group properties of these transformations, an additional rotation of the coordinate axes of the frames K and K" appears at the Thomas-Wigner angle Ω, depending on $\vec{v}_1$ and $\vec{v}_2$. It is quite important that such a rotation occurs in measured space-time coordinates, *i.e.*, it can be really detected. It defines a principal possibility to experimentally distinguish the hypotheses **A**=**L** and **A**≠**L**. We also notice that for collinear $\vec{v}_1$ and $\vec{v}_2$, Ω=0, and the absolute velocity is not observable. This result corresponds to the second implication of GRP, obtained in section 2. Hence, in corresponding experiments, testing CETs, these velocities should be non-collinear.

Among admissible space-time theories that assume **A**≠**L**, the simplest case corresponds to the choice **A**=**G**, where **G** is the matrix of Galilean transformation: $G_{aa}=1$, $G_{i0}=-v_i$, and all others $G_{ab}=0$. Substituting matrix **G** in place of matrix **A** in Eqs. (27), one gets the following coefficients of matrix **B**:

$$B_{00} = g, \; B_{i0} = 0, \; B_{0i} = \frac{v_i}{c^2}g, \; B_{ij} = d_{ij}\frac{v_i v_j}{v^2}(1 - 1/g) \qquad (33)$$

where $d_{ij}$ is the Kronekker symbol. Further substitution of Eqs. (33) into Eq (9) allows one to determine a dependence of physical space-time four-vectors on the absolute velocity $\vec{v}$ of some arbitrary inertial reference frame K:

$$\vec{r}_{ph}(\vec{v}) = \vec{r}_{ph}(v=0) + \frac{\vec{v}(\vec{r}_{ph}(v=0),\vec{v})}{v^2}\left[\sqrt{1-(v^2/c^2)} - 1\right], \tag{34}$$

$$t_{ph}(\vec{v}) = \frac{t_{ph}(v=0)}{\sqrt{1-(v^2/c^2)}} + \frac{\vec{r}_{ph}(v=0)\vec{v}}{c^2\sqrt{1-(v^2/c^2)}}. \tag{35}$$

For the time interval in a fixed spatial point of the frame K ($r_{ph}=0$), we obtain the dependence of $t_{ph}$ on $\vec{v}$ (see, Eq. (35)):

$$t_{ph}(\vec{v}) = t_{ph}(v=0)/\sqrt{1-(v^2/c^2)}, \tag{36}$$

that means an absolute dilation of time by factor $\sqrt{1-(v^2/c^2)}$. Furthermore, one obtains from Eq. (34):

$$(\vec{r}_{ph}(\vec{v}),\vec{v}) = (\vec{r}_{ph}(v=0),\vec{v})\sqrt{1-v^2/c^2}, \quad [\vec{r}_{ph}(\vec{v})\times\vec{v}] = [\vec{r}_{ph}(v=0)\times\vec{v}], \tag{37}$$

that means an absolute contraction of moving scale along a vector of absolute velocity by factor $\sqrt{1-(v^2/c^2)}$ (Fitzgerald-Lorentz hypothesis). Finally, transformation (1) (under **A**=**G**)

$$(x_{ph})_a = [G_{ab}(\vec{v}_1 - \vec{v}_2)](x''_{ph})^b$$

leads to the Galilean law of speed addition for the physical light velocity $c_{ph}$.

Thus, we have got a full set of the Lorentz ether postulates in case **A**=**G**.[2] However, the physical space-time in the Lorentz ether theory is not observable in an arbitrary inertial reference frame, while the measured four-vectors $x_{ex}$ obey the Lorentz transformations in the form of (31). (This important circumstance concerning a difference of physical and measured four-vectors for oblique-angled metrics of space-time was dropped by Lorentz and his successors). Therefore, we may consider the Lorentz ether theory (LET) as one of the CETs defined above, and the simplest among them. Due to this fact, the application of the Lorentz ether postulates for the explanation of "null" results of all experiments searching for "ether wind speed" was always successful. At the same time, now we get a possibility to proceed not only from the Lorentz ether postulates, but from the complete kinematics of LET. Its full description is given by the following equations, obtained above:

$$(x'_{ph})_a \neq (x'_{ex})_a \text{ (for the absolute frame)}, \tag{38}$$

$$(x_{ex})_a = L_{ab}(\vec{v}_1)[L^{-1}(\vec{v}_2)]^{bg}(x''_{ex})_g \tag{39}$$

$$(x_{ph})_a = [G_{ab}(\vec{v}_1 - \vec{v}_2)](x''_{ph})^b \tag{40}$$

$$\vec{r}_{ph}(\vec{v}) = \vec{r}_{ph}(v=0) + \frac{\vec{v}(\vec{r}_{ph}(v=0),\vec{v})}{v^2}\left[\sqrt{1-(v^2/c^2)} - 1\right], \tag{41}$$

$$t_{ph}(\vec{v}) = \frac{t_{ph}(v=0)}{\sqrt{1-(v^2/c^2)}} + \frac{\vec{r}_{ph}(v=0)\vec{v}}{c^2\sqrt{1-(v^2/c^2)}}. \tag{42}$$

One should notice that the Galilean transformation itself does not restrict a value of limited velocity, it can be infinite. However, in case of LET such a restriction is established by the equality $(x'_{ph})_a \nLeftrightarrow (x'_L)_a$, which simultaneously means that the Galilean transformations in LET are "admissible", and they act within pseudo-Euclidean geometry.

---

[2] Let us recall the postulates of Lorentz ether theory in its modern form:
**1)** There is an absolute reference frame $K_0$, wherein light velocity is isotropic and equal to $c$. **2)** In an arbitrary reference frame K, moving at constant velocity $\vec{v}$ in $K_0$, the velocity of light is equal to $\vec{c}' = \vec{c} - \vec{v}$. **3)** In this reference frame K time is dilated by $\sqrt{1-v^2/c^2}$ times. **4)** In this reference frame K a linear scale is contracted by $\sqrt{1-v^2/c^2}$ times along the vector $\vec{v}$.

The most important physical consequence of kinematics of LET, expressed by Eqs. (38)-(42), is a principal possibility to detect experimentally the absolute velocity $\vec{v}$ of an inertial reference frame, due to the mentioned above dependence of the Thomas-Wigner angle $\Omega$ on $\vec{v}$. It can be seen in the problem of diametrical synchronization of distant clocks by a moving rod.

Let two clocks $Cl_1$ and $Cl_2$ be placed upon the $x$ axis of some inertial reference frame K at rest in the absolute frame $K_0$. The distance between $Cl_1$ and $Cl_2$ is equal to $L$. Let some rod with a proper length $L$ moves along the $y$ axis at a constant velocity $\vec{u}$. The axis of the rod is parallel to the $x$ axis, and the coordinates of its opposite ends upon the $x$ axis coincide with the respective coordinates of $Cl_1$ and $Cl_2$. So at the instant when the rod is intersecting the axis $x$, it is simultaneously touching the $Cl_1$ and $Cl_2$ in the frames K and $K_0$. We assume that at the touch moment each clock emits a short light pulse towards to the time analyzer (TA) placed between the clocks. Thus, when K rests in $K_0$, the indication of TA is $\Delta t = 0$.

Now consider the same problem when the frame K moves at the constant absolute velocity $\vec{v}$ along the $x$ axis (see, Fig. 3). One requires to find in the laboratory frame K an indication $\Delta t$ of TA.

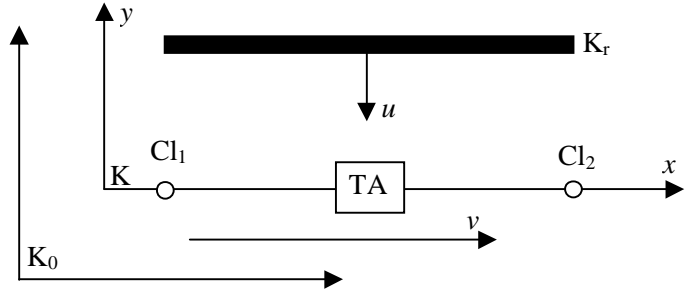

Fig. 3. The scheme of "diametrical" synchronization of distant clocks by moving "ideal" rod.

We attach the frame $K_r$ with the moving rod. The velocity of $K_r$ with respect to K is equal to $u$ along the $y$ axis, while the velocity of K with respect to $K_0$ is equal to $v$ along the axis $x$. In order to calculate the value $\Delta t$ within SRT, one should apply a special Lorentz transformation from $K_r$ to K. Hence, we get $\Delta t = 0$, as in the case $v=0$.

Let us calculate the value $\Delta t$ for Fig. 3 in the Lorentz ether theory ($\mathbf{A}=\mathbf{G}$). In such a case, the measured space-time coordinates $x_{ex}$ are subjected to the transformation (31), according to which we should apply the succession $K_r \to K_0 \to K$ with the velocities $\vec{V} = \vec{v} \oplus \vec{u}$ and $\vec{v}$, respectively. Such a transformation for the $x_{ex}$ coordinates entails a relative rotation of $K_r$ and K coordinate axes at the angle [1]

$$\Omega \approx uv/2c^2 \ . \tag{43}$$

Hence, at the instant when the left end of rod touches the $Cl_1$, its right end has a non-zero coordinate $\Omega L \approx Luv/2c^2$ upon the axis $y$. From there

$$\Delta t \approx \Omega L/u = Lv/2c^2 \ . \tag{44}$$

Since the measured light velocity is isotropic, that Eq. (44) directly gives the indication of TA. However, Eq. (44) has no physical interpretation in the measured space-time coordinates. On the contrary, this equation has a clear physical meaning in physical space-time. Indeed, here we get an absolute contraction of moving rod along its resultant absolute velocity $\vec{V} = \vec{v} \oplus \vec{u}$. According to Eq. (37), the projection of the rod perpendicular to $\vec{V}$ remains un-changed. Let us denote it as $L\sin a$, where $a$ is the angle between $\vec{L}$ and $\vec{V}$. A projection of the rod which is parallel to $\vec{V}$ becomes equal to $L\sqrt{1-V^2/c^2}\cos a$. As a result, the axis of the rod turns out with respect to the axis $x$ at the angle $j \approx uv/2c^2$ in comparison with the case $v=0$ (to order of approximation $c^{-2}$). Further, the physical light velocity along the $x$ axis of the laboratory frame K is equal to $c_+ = c - v$, and in the opposite direction $c_- = c + v$. Hence, the indication of TA is

$$\Delta t \approx \frac{jL}{u} + \frac{L}{2(c+v)} - \frac{L}{2(c-v)} \approx \frac{Lv}{2c^2}.$$

This coincides with Eq. (44). Thus, Eq. (44) can be interpreted as the real appearance of the properties of physical space-time, in spite of the impossibility to directly measure $x_{ph}$. In particular, the calculations presented allow one to consider Eq. (44) as an inference of the absolute contraction of rod as well the anisotropy of physical light speed $c_{ph}$ in the moving laboratory frame K. From a formal viewpoint, such a result follows from the dependence of $\Omega$ on $\vec{v}$ in experimentally measured coordinates $x_{ex}$, caused by the general transformation rule (39). Thus, a formal application of the transformation (39) for Minkowskian four-vectors $x_L$ (leading to the measurable dependence of $\Omega$ on $\vec{v}$) finds a physical interpretation in the $x_{ph}$ coordinates, despite the impossibility of observing the $x_{ph}$ four-vectors experimentally.

One can finally add that the solution (44) could also be produced by conventional relativistic calculations when the relative velocity of K and $K_0$ is equal to $\vec{v}$, while the relative velocity of $K_r$ and $K_0$ is equal to $\vec{V} = \vec{v} \oplus \vec{u}$. However, such a motion diagram of the frames $K_r$, K in $K_0$ differs from the motion diagram in Fig. 3. Thus, the solution (44) and the solution $\Delta t = 0$ within SRT correspond to different physical problems, while within LET these solutions reflect a dependence of $\Omega$ on $\vec{v}$ for a fixed experimental instrument ($Cl_1$, $Cl_2$ + moving rod).

*4.1. Notes about classical dynamics and classical electrodynamics of CETs*

Now we make a further step to generalization of CETs, stating that physical and measured values should be inevitably distinguished not only for space-time four-vectors, but for any functions of $t$, $\vec{r}$. Then the quantities, depending on $x_{ex}$ and constituting measured four-vectors (energy-momentum, charge density-current density, etc.), obey the Lorentz transformations, while their corresponding physical magnitudes (depending on $x_{ph}$) are subjected to the transformation **A**. This result follows from the general transformation rule for a four-vector $X$:

$$X^i = (\partial x^i / \partial x'^k) X'^k.$$

Indeed, one can see from Eq. (1) that $(\partial x_{ph})^i / (\partial x'_{ph})^k = A^i_k$, and $(X_{ph})^i = A^i_k (X'_{ph})^k$, or $(X_{ph})_i = A_{ik}(X'_{ph})^k$, which coincides with Eq. (1).

In the case **A**=**G** (LET) we get the Galilean transformation for $(X_{ph})_i$. In particular, when $X_{ph} = (E_{ph}, \vec{p}_{ph})$, where $E_{ph}$ is the physical energy, and $\vec{p}_{ph}$ is the physical momentum, we obtain

$$\vec{p}_{ph} = \vec{p}'_{ph} - \frac{\vec{v}E'_{ph}}{c^2}, \quad E_{ph} = E'_{ph}. \tag{45}$$

Then a dependence of physical momentum and energy on the absolute velocity $\vec{v}$ is determined similarly to Eqs. (34), (35):

$$\vec{p}_{ph}(\vec{v}) = \vec{p}_{ph}(v=0) + \frac{\vec{v}(\vec{p}_{ph}(v=0) \cdot \vec{v})}{v^2}\left[\sqrt{1-(v^2/c^2)} - 1\right], \tag{46}$$

$$E_{ph}(\vec{v}) = \frac{E_{ph}(v=0)}{\sqrt{1-(v^2/c^2)}} + \frac{\vec{p}_{ph}(v=0) \cdot \vec{v}}{\sqrt{1-(v^2/c^2)}}. \tag{47}$$

Note that in the rest frame of particle $\vec{p}_{ph} = 0$, and $E_{ph}(\vec{v}) = E_{ph}(v=0)/\sqrt{1-(v^2/c^2)}$. Taking into account that $E_{ph}(v=0) = mc^2$ ($m$ is the rest mass of particle), we get $E_{ph}(\vec{v}) = mc^2/\sqrt{1-(v^2/c^2)}$, that is an usual expression of relativistic physics, but with alternative physical interpretation: a total physical energy of moving particle (and, correspondingly, total physical mass) is fully determined by its absolute velocity. In contrast, a measured energy and measured mass depend on a relative velocity solely.

The scalar $j$ and vector $\vec{A}$ potentials, as the functions of space and time coordinates, also have the physical and measured values in CETs. The same is true for the electric $\vec{E}$ and magnetic $\vec{B}$ fields. The measured fields $\vec{E}_{ex}(t_{ex}, \vec{r}_{ex})$, $\vec{B}_{ex}(t_{ex}, \vec{r}_{ex})$ obey the conventional Maxwell's equations in any inertial reference frame. Physical electromagnetic fields $\vec{E}_{ph}(t_{ph}, \vec{r}_{ph})$ and $\vec{B}_{ph}(t_{ph}, \vec{r}_{ph})$ obey the Maxwell's equations only in the absolute frame. Concerning these equations in an arbitrary frame of references, we mention that an absolute motion of an inertial frame in CETs, in fact, induces its admissible coordinate transformation, depending on the absolute velocity. We stress that such a coordinate transformation in CETs is an objective property of nature, on the contrary to purely mathematical coordinate transformations in SRT. That is why CETs leads to alternative to SRT physics. At the same time, a mathematical identity of the admissible coordinate transformations in two these theories allows applying the well-developed coordinate formalism of SRT to physical problems of CETs. For this purpose we have to determine the metric tensor $g_{ph}$ in physical space-time of moving inertial frame, where the physical space and time intervals obey the transformation **A** in Eq. (1). In order to solve this problem, we use the expression for space-time interval:

$$s^2 = (g_{ph})_{ab} x_{ph}^{\,a} x_{ph}^{\,b} = (g_{ph})_{ab}\left(B_g^a x_L^{\,g}\right)\left(B_l^b x_L^{\,l}\right) = (g_L)_{gl} x_L^{\,g} x_L^{\,l},$$

where $g_L$ stands for the Minkowskian metric tensor. From there

$$(g_{ph})_{ab} B_g^a B_l^b = (g_L)_{gl}. \tag{48}$$

The latter expression represents a system of ten linear equations with respect to ten independent parameters $(g_{ph})_{ab}$ (a total number of coefficients $g_{ab}$ is equal to 16, but $g_{ab} = g_{ba}$, and only ten independent parameters remain). A solution of this system is:

$$(g_{ph})_{ab} = (B^{-1})_{ga}^{\,}(B^{-1})_b^{\,g}, \tag{49}$$

where the matrix **B** is defined by Eqs. (27). For physical space-time transformation **A**=**G** (LET), the coefficients of matrix **B** are found from Eqs. (33). Then the straightforward calculations give:

$$g_{00} = 1 - v^2/c^2, \quad g_{0i} = g_{i0} = v_i/c, \quad g_{ii} = -1, \text{ and others } g_{ij} = 0. \tag{50}$$

By the way, we see from Eqs. (50) that for $g_{00} > 0$ (the requirement of GRP), the absolute velocity *v* is always less than *c*. This again signifies that we work within admissible Galilean transformations, and do not leave pseudo-Euclidean geometry of empty space-time. Substituting the coefficients (50) into the Maxwell equations, written in arbitrary admissible coordinates (see, e.g. [1, 3]), we can find their form in physical space-time of LET. Within the same coordinate formalism we can derive a law of transformation of physical electric and magnetic fields, using the metric tensor (50). This is a quite formal mathematical task. However, a physical analysis of classical electrodynamics in LET, written in physical space-time with the metric tensor (50), represents a very extensive problem. At the same time, it is important to remember that using measured space-time coordinates, we always get the corresponding "measuring" functions of space-time, and observable world looks almost the same, like in conventional relativistic dynamics and classical electrodynamics. The difference appears in successive space-time transformations with non-collinear relative velocities. This problem in its application to dynamics and electrodynamics will be analyzed elsewhere. However, without such a detailed analysis, we already can notice that the conception about "physical" and "measured" quantities appears to be useful in resolution of a number of relativistic paradoxes of classical electrodynamics. One of them is presented in Fig.4. Let there be two charges point-like particles with the charge +*q* and –*q*, respectively, being inside a hollow neutral tube. The tube is placed into a parallel plate condenser, creating a homogeneous electric field *E* along the axis *y*. The tube has a single degree of freedom to move along the axis *y*. A gravitation field is absent, so that the masses of particles are not relevant. Initially both particles are fixed inside the tube and they rest with respect to each other and the condenser. The resultant force acting on the tube along the axis *y* is composed as the sum of *qE* and –*qE*, and equal to zero. Now we imagine that the positively charged particle (p) can move inside the tube, and it acquires a constant velocity *v* in the negative *x*-direction (Fig. 4). We want to compute the force acting on the tube along the axis *y*.

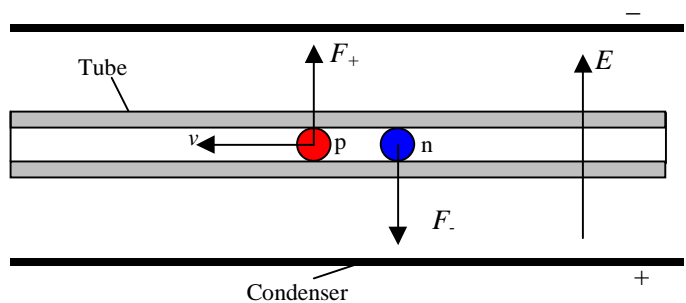

Fig. 4. The oppositely charged point-like particles "p" and "n" are placed into a hollow tube. All system is in the inner volume of the parallel plate condenser, which creates the electric field *E* along the axis *y*. The particle "n" is fixed inside the tube and rests with respect to condenser. The particle "p" moves at the constant velocity *v* along the axis -*x*. Here we assume a presence of an external mechanical force along the axis *x*, which compensates an electrical attraction of the particles. We also exclude a weak electric interaction between the condenser and the thin wall tube due to its electric polarization.

It seems that the problem is trivial: the magnetic field is absent in the rest frame of condenser ($K_C$), and the total force acting on the moving particle "p" is directed along the axis *y* and equal to $F^+=qE$, like in the case *v*=0. Since the force acting on the particle "n" is $F^-=-qE$, that the resultant force, exerted on the tube due to two charged particles, is equal to zero.

A paradox appears, when we compute the same force in the rest frame $K_p$ of particle "p" within SRT. According to the relativistic force transformation law [1]

$$\vec{F}' = \frac{\sqrt{1-v^2/c^2}\,\vec{F} + \dfrac{\vec{v}\cdot(\vec{v}\cdot\vec{F})\left(1-\sqrt{1-v^2/c^2}\right)}{v^2} + \dfrac{\vec{v}\cdot(\vec{u}\cdot\vec{F})}{c^2}}{1+(\vec{v}\cdot\vec{u})/c^2}, \qquad (51)$$

($\vec{v}$ is the velocity of the frame K in the frame K', and $\vec{u}$ is the velocity of particle in the frame K), the particle "p" in its rest frame experiences the force

$$F'^{+}{}_y = F^{+}{}_y\big/\sqrt{1-v^2/c^2} = qE\big/\sqrt{1-v^2/c^2}. \qquad (52)$$

Computing the force acting on the particle "n" in this frame $K_p$, we have to accomplish a reverse force transformation from $K_p$ to $K_C$, taking into account that in the frame $K_C$, $F^{-}=-qE$. Then we obtain from Eq. (51):

$$F'^{-}{}_y = F^{-}{}_y\sqrt{1-v^2/c^2} = -qE\sqrt{1-v^2/c^2}, \qquad (53)$$

and $F'^{+}{}_y \neq F'^{-}{}_y$.

Let us show that Eqs. (52) and (53) are in agreement with a direct calculation of the Lorentz forces acting on both particles in the frame $K_p$. Indeed, according to the relativistic transformation of electromagnetic field, the moving in $K_p$ condenser produces the electric field $E'_y = E\big/\sqrt{1-v^2/c^2}$ and the magnetic field $B'_z = vE\big/c^2\sqrt{1-v^2/c^2}$ as well. Hence, the force acting on the particle "p" is $F'^{+}{}_y = qE'_y = qE\big/\sqrt{1-v^2/c^2}$, which coincides with Eq. (52). The force experienced by the moving particle "n" is written in accordance with the Lorentz force law $\vec{F} = q(\vec{E}+\vec{v}\times\vec{B})$ as

$$F'^{-}{}_y = -qE'_y - qvB'_z = -\frac{qE}{\sqrt{1-v^2/c^2}} - \frac{qv^2 E}{c^2\sqrt{1-v^2/c^2}} = -qE\sqrt{1-v^2/c^2},$$

which coincides with Eq. (53). Therefore, the resultant force, acting on the tube along the axis $y$ in the frame $K_p$ is determined as the sum of time-independent forces (52) and (53), and it is not vanishing:

$$F'_{\Sigma y} = F'^{+}{}_y + F'^{-}{}_y = \frac{qE}{\sqrt{1-v^2/c^2}} - qE\sqrt{1-v^2/c^2} = \frac{qEv^2}{c^2\sqrt{1-v^2/c^2}} \neq 0. \qquad (54)$$

Thus, we derive the paradoxical result: in the frame $K_p$ the tube acquires the acceleration along the axis $y$ due to the force (54), while in the rest frame of condenser it should remain at rest.

The paradox can be resolved in CETs, using the "physical" forces, for example, for LET. Then we can get from Eq. (45)

$$\frac{d\vec{p}_{ph}}{dt_{ph}} = \frac{d\vec{p}'_{ph}}{dt'_{ph}} - \frac{\vec{v}\,dE'_{ph}}{c^2 dt'_{ph}}.$$

where we applied the Galilean transformation of physical time, $t_{ph} = t'_{ph}$. Denoting the force $\vec{F} = d\vec{p}/dt$, and taking into account that $dE'_{ph}/dt'_{ph} = \vec{w}\cdot\vec{F}'_{ph}$ ($\vec{w}$ is the absolute velocity of particle), we obtain:

$$\vec{F}_{ph} = \vec{F}'_{ph} - \frac{\vec{v}\cdot(\vec{w}\cdot\vec{F}'_{ph})}{c^2}. \qquad (55)$$

Regardless of specification of the absolute frame $K_0$ for the problem in Fig. 4, we have $\vec{w}\perp\vec{F}'_{ph}$, and $\vec{F}_{ph} = \vec{F}'_{ph}$. Therefore, the charged particles "p" and "n" experience the force $qE_y$ and $-qE_y$, correspondingly, and the resultant force is vanishing for any inertial frame of observations. It resolves the paradox.

## 5. Possible experimental tests of CETs

As we concluded above, the difference between SRT (**A=L**) and CETs (**A≠L**) appears at the experimental level only in successive space-time transformations with non-collinear relative velocities. Such is a property of space-time transformations in CETs (Eqs. (31), (32)), which corresponds to the general conclusions of Section 2. It follows from there, that an instrument for measuring an absolute velocity must contain moving inertial parts, in order to deal with such transformations. Then a general idea of an experiment for a choice between SRT and CETs can be described with the help of the diagram in Fig. 5. It shows the absolute frame $K_0$, laboratory frame K (moving at the constant absolute velocity $\vec{v}$) and frame $K_i$, attached to some moving inertial part of a measuring instrument in K.

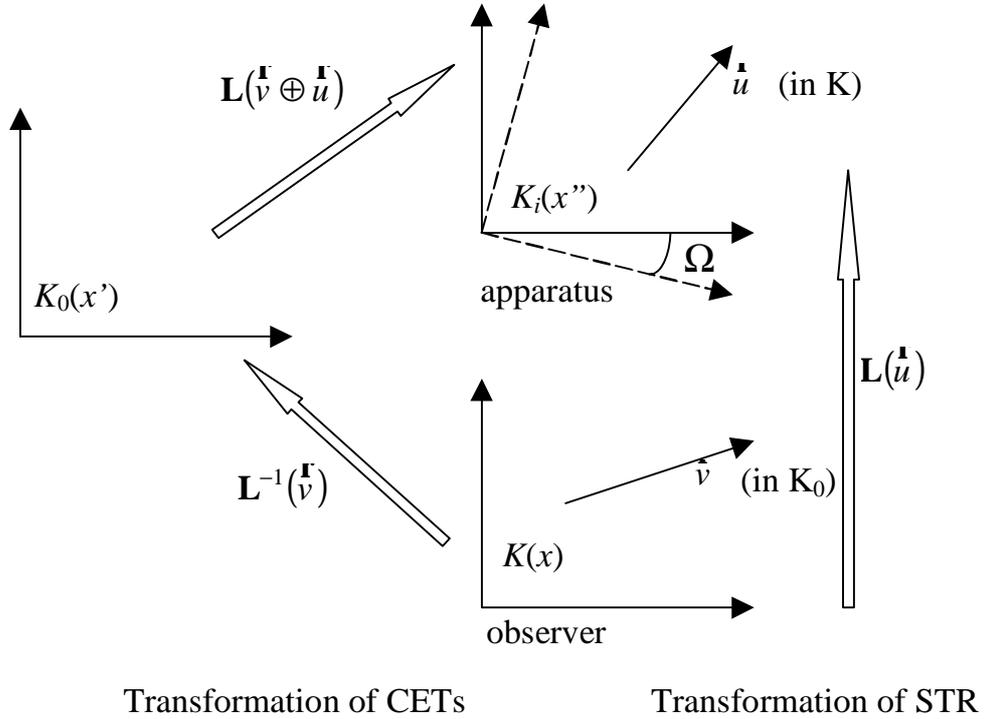

Fig. 5. General idea for an experiment to provide an experimental test of covariant ether theories.

In our laboratory we specify a constant velocity $\vec{u}$ of $K_i$ in K. In such a case for the hypothesis **A=L** we apply a direct rotation-free Lorentz transformation K→$K_i$ for calculation of the reading of the measuring device. Hence, according to SRT we get a vanishing value of absolute velocity. Under the hypothesis **A≠L**, Nature does not "know" a direct rotation-free Lorentz transformation between K and $K_i$, and "operates" with the absolute velocities of these frames $\vec{v}$ and $\vec{v} \oplus \vec{u}$. Hence, in order to calculate an indication of the measuring device, we must apply the successive Lorentz transformations K→$K_0$→$K_i$ according to Eq. (31). (A direct Lorentz transformation from K to $K_i$ is also possible, but it will not be rotation-free). In this case the axes of the frames K and $K_i$ are turned out at the Thomas-Wigner angle $\Omega$, that, in principle, changes the state of the measuring instrument. Since $\Omega$ depends on the absolute velocity $\vec{v}$ of the laboratory frame K, the state of the measuring instrument will depend on $\vec{v}$, too. There is only one particular case ($\vec{v}$ is collinear to $\vec{u}$) where $\Omega = 0$, and the state of the measuring instrument has to be unchanged for any magnitude of absolute velocity of the labo-ratory frame. Therefore, all experiments searching for "ether wind" velocity with experimental instruments containing moving inertial parts, aiming to measure non-relativistic effects under collinear $\vec{v}$ and $\vec{u}$ (see, *e.g.*, [10-12]), in fact checked the GRP, not the Einstein relativity principle.

Thus, an experiment for testing of CETs must contain moving inertial parts with non-collinear velocities $\vec{v}$ and $\vec{u}$, and be aimed to measure a dependence of the angle Ω on the absolute velocity of the laboratory frame. To the order of magnitude $c^{-2}$, and for orthogonal vectors of $\vec{v}$ and $\vec{u}$, this dependence is defined by Eq. (43). A direct measurement of this dependence in a laboratory-scale experiment is impractical. Indeed, the absolute speed $v$ could be taken as about $10^{-3}c$ (typical velocities of Galaxy objects). The maximum value of $u$ could be about $10^3$ m/s. Hence, the angle Ω takes on the value $3\times10^{-9}$, *i.e.*, well below any limit of practicability in a laboratory experiment.

An analysis of possible experimental schemes for indirect measurement of the angle Ω can be greatly simplified through numerical estimation of eventual non-relativistic effects, proceeding from their dimension. The experiments, looking for a change of length associated with $\Omega(\vec{v})$ dependence, give an effect of the order of magnitude $L\Omega$. Here $L$ is some length, which is equal to about 1 m in the laboratory scale. In such a case we get $L\Omega \approx 3\times10^{-9}$ m, which is impossible to measure in practice. A corresponding change of time has a dimension $L\Omega/u \approx Lv/c^2 \approx 3$ ps, a time interval within the range of present technology, but not for the mechanical parts necessarily involved. Further, one can rearrange the experiment into a "speed experiment" looking at the term $\Omega u$, and the latter arrangement could be further transformed into frequency measurement via the Doppler effect ($\Omega u/c$). For the last case one is looking at the term $u^2v/c^3$, the latter being about $10^{-14}\ldots10^{-15}$ (for $u=10^2\ldots10^3$ m/s) - a value accessible practically conveniently by the Mössbauer effect, at least as far as a laboratory-scale experiment is concerned. One of such experiments is considered in sub-section 6.1. Finally, one may consider an electromagnetic experiment, where the eventual non-relativistic effect should be $V\Omega$ ($V$ is the voltage). For example, for $V=10^4$ V and $\Omega \approx 10^{-9}$, one gets $V\Omega=10^{-5}$ V. This value is acceptable for laboratory measurements. However, the analysis of such experiments requires to develop classical electromagnetic theory of CETs in detail, which can be done elsewhere.

*5.1. Proposed experiments for test of CETs, based on the Mössbauer effect*

One should mention that the Mössbauer experiments for test of CETs, in their comparison with known experiment by Champeney et al. [10], should be more sensitive by $u/c$ times. In fact, only past decade gave an opportunity to realize such high-sensitive experiments, when the technique for resonant synchrotron radiation and for resonant detection of gamma-quanta had been developed. Possible schemes of such experiments with the resonant synchrotron radiation have been described in ref. [13] (for $^{67}$Zn isotope) and in ref. [14] (for $^{57}$Fe isotope). In this paper we propose another experiment, which uses a radioisotope Mössbauer spectroscopy with resonant detection of gamma-quanta of $^{119}$Sn.

Let us consider the following experimental scheme. Let there be two rotors with the equal radius $r$, lying on the same plane and separated by the distance $L$. Under synchronic rotation at the angular frequency $w$ these rotors drive a rod with the length $L$, as shown in Fig. 6. The Mössbauer source S and receiver R are fixed on the opposite sides of the rod. In this geometry we measure a relative Doppler shift between emission and absorption lines, which, as we will see below, is a function of the "absolute" velocity of Earth $\vec{v}$. In order to obtain this function in the explicit form, let us consider a diagram of velocities of source and receiver in the absolute frame (Fig. 7). In this diagram the vector of tangential velocity rotates clockwise at the angular frequency *w* for both the source and receiver. For simplicity we can imagine these vectors as some "clock arrows", and within LET, where the Galilean transformations are valid, the physical directions of both " clock arrows" coincide at any fixed instant. (A similar diagram drawing according to SRT, gives a corresponding retardation for the right "clock arrow"). However, during a time of light propagation from the first clock to the second clock, the right "arrow" has time to turn out at some angle $\Delta j$, which causes a corresponding linear Doppler shift between emission and absorption lines. It is obvious that the time of propagation of gamma-quanta from S to R (and hence, the value of $\Delta j$) depends on the angle between the vector of "absolute" velocity $\vec{v}$ and the line S-R. Therefore, the same dependence should be detected, while measuring the relative energy shift between the source's and receiver's resonant lines.

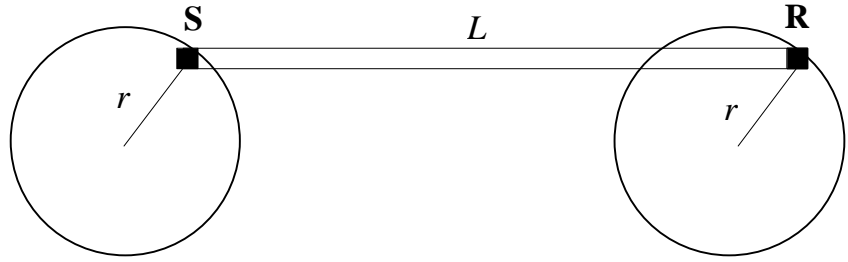

Fig. 6. Schematic of the experiment for measurement of an absolute velocity

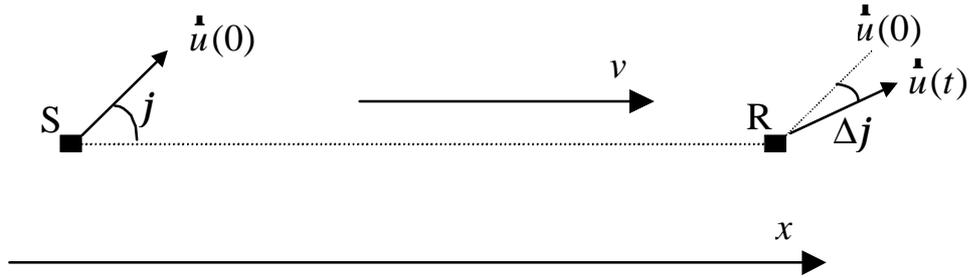

Fig. 7. Diagram of the velocities of source and receiver in the Mössbauer experiment of Fig. 6.

Such is a general idea of the experiment, which should be confirmed by calculations. First we adopt for simplicity that the vector $\vec{v}$ is collinear to the line S-R (Fig. 7). Let us write an expression for the frequency of emitted radiation $n_{em}$, considering the process from the absolute frame and using known equation for the Doppler effect:

$$n_{em} = \frac{n_0 \sqrt{1-(V^2/c^2)}}{1-(\vec{V}\vec{n}/c)}, \qquad (56)$$

where $\nu_0$ is the proper frequency of Mössbauer radiation, $\vec{V}$ is the velocity of source in the absolute frame $K_0$ at the emission instant of gamma-quanta, and $\vec{n}$ is the unit vector in the emitting direction. An expression for the absorption frequency $\nu_a$ takes the form:

$$n_a = \frac{n_{em}[1-(\vec{V}'\vec{n}/c)]}{\sqrt{1-(V'^2/c^2)}} = \frac{n_0\sqrt{1-(V^2/c^2)}[1-(\vec{V}'\vec{n}/c)]}{[1-(\vec{V}\vec{n}/c)]\sqrt{1-(V'^2/c^2)}}, \tag{57}$$

where $\vec{V}'$ is the velocity of receiver in the absolute frame at the absorption instant of gamma-quanta. Further, the Einstein law of speed composition, applied to the absolute frame, gives

$$V_x = \frac{u\cos j + v}{1+(uv\cos j/c^2)}; V_y = \frac{u\sin j\sqrt{1-(v^2/c^2)}}{1+(uv\cos j/c^2)};$$

$$V'_x = \frac{u\cos(j+\Delta j)}{1+(uv\cos(j+\Delta j)/c^2)}; V'_y = \frac{u\sin(j+\Delta j)\sqrt{1-(v^2/c^2)}}{1+(uv\cos(j+\Delta j)/c^2)}, \tag{58}$$

where $u$ stands for the linear velocity of the source and receiver upon the rotor in the laboratory frame, $j$ is the angle of the vector $\vec{u}$ with the axis $x$ at the moment of emission of gamma-quanta by the source. We can get the components of the vector $\vec{n}$ from the following expressions (to the accuracy $c^{-2}$, sufficient for further calculations):

$$n_x = 1 - \frac{u^2\sin^2 j}{2c^2}; n_y = \frac{u\sin j}{c} - \frac{uv\sin j}{2c^2} - \frac{u^2\sin j \cos j}{2c^2}. \tag{59}$$

Inserting corresponding values from Eqs. (58) and (59) into Eq. (57), we get after simple manipulations to the order $c^{-3}$:

$$n_a = n_0\left(1 - \frac{u\Delta j \sin j}{c}\right) \tag{60}$$

Here $\Delta j = w\Delta t = wL/(c-v)$ ($\Delta t$ is the time of gamma-quanta propagation from the source to receiver), $j = wt$. Substituting the value of $\Delta j$ into Eq. (60), and decomposing into series the obtained expression to the order $c^{-3}$, we obtain

$$n_a = n_0\left(1 - \frac{uwL\sin wt}{c^2} - \frac{uvwL\sin wt}{c^3}\right) \tag{61}$$

Eq. (61) describes the energy shift between emission and absorption lines. The second term in brackets of *rhs* of Eq. (61) has a meaning of "gravitational shift" of gamma-quanta frequency in the non-inertial reference frame attached to the source and receiver of Mössbauer radiation. Indeed, in accordance with the equivalence principle, an oscillating gravitational field with the potential $uw\sin wt$ appears in this frame. And the measured energy shift of gamma-quanta depends on the potential difference between S and R, according to known result of general relativity theory. Such a shift is vanishing for $j = 0$, i.e., for the case when the acceleration is perpendicular to the line S-R.

The most interesting is the third term in brackets of Eq. (61), which is proportional to the "absolute" velocity $v$. One can see that for $j = 0$ (i.e., when the momentary velocities of the source and receiver are collinear to $\vec{v}$), this term becomes to be equal to zero. Such a result directly reflects the fact that under addition of Lorentz boosts with the collinear velocities, the Lorentz operators commutate with each other and an absolute velocity is not observable in CETs. We also notice that this term is proportional to $c^{-3}$, i.e., its value is greatly less than in Champeney et al. experiment, where the eventual term was proportional to $c^{-2}$. As we mentioned earlier, it leads to stronger requirements to the energy sensitivity of the considered experiment in comparison with the Champeney et al. experiment. Let us estimate numerically the third order term in Eq. (61) for $\sin j = 1$. We adopt $L=1$ m, the rotation frequency $v=200$ Hz ($w \approx 1250$ Hz); the rotor's radius $r=10$ cm ($u=wr=125$ m/s); $v/c=10^{-3}$. Then the relative energy shift is equal to $((n_a - n_0)/n_0 = 1,7\times 10^{-15}$. In order to measure such an energy shift, we suppose to use a single line resonant scintillation detector (RSD) [15], as a receiver R, with a combination of single line Mössbauer source S. Such detectors have been developed to reach a sufficient progress for $^{119}$Sn (23.9 keV) and other isotopes of Mössbauer spectroscopy. It consists of a thin scintillating organic film, where the absorbing medium (converter) is distributed, and a photomultiplier. The converter transforms a beam of gamma-quanta into conversion electrons accompanying decay of resonantly exited nuclei. Due to the very low gamma-counting efficiency by the organic film and the high counting efficiency for low-energy conversion electrons such a detector has a high selectivity value $S=\eta_0/\eta_\infty$ ($\eta_0$ is the counting efficiency for resonant gamma-quanta, while $\eta_\infty$ is the counting efficiency for non-resonant gamma-quanta). For RSD the value $S>50$ [15]. Hence, one may expect a great increase in sensitivity of RSD to the relative energy shift in comparison with ordinary detectors of gamma-quanta, applied in Champeney et al. experiment.

Moreover, on the contrary to the Champeney et al. experiment, an RSD allows us to realize a continuous registration of gamma-quanta as a function of a rotating angle $wt$ due to application of special system of light-guides between the resonant scintillator and photomultiplier (Fig. 8). Such an arrangement allows one to look for the harmonic oscillations in measured signal ($\sin wt$ in Eq. (61)) and thus, to exclude an influence of possible time drift of the detector parameters as well as possible material deformations in the rotating frame.

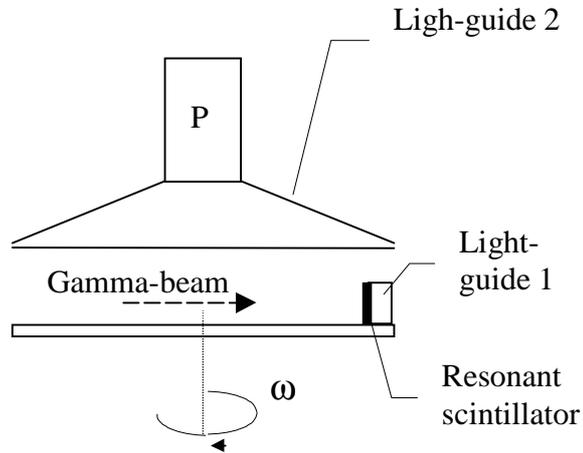

Fig. 8. An optical coupling of the resonant scintillator with a photomultiplier P. S is a Mössbauer source.

## 6. Conclusions

1) Consideration of all hypothetical ether theories of empty space-time, admitting pseudo-Euclidean geometry with oblique-angled metrics in arbitrary inertial frame, should be based on distinguishing between physical and measured space-time four-vectors. General analysis of the properties of admissible space-time transformations shows that in any theory adopting the general relativity principle and symmetries of space-time, the measured space and time intervals always obey the Lorentz transformation, regardless of a concrete choice of physical space-time transformation. The latter circumstance makes it possible to explain all known experimental results in physics of empty space-time within an infinite number of admissible space-time theories, called "covariant ether theories".

2) SRT is unique among admissible theories of empty space-time because it directly asserts the equality between measured and physical four-vectors under optimal measurements. Adoption of such an equality defines the possibility of direct rotation-free Lorentz transforma-tion between two arbitrary inertial frames. This is impossible in all other admissible space-time theories, named as covariant ether theories. An absolute motion of inertial frame in CETs, in fact, induces its admissible coordinate transformation, depending on an absolute velocity. We stress that such a coordinate transformation in CETs is an objective property of nature, on the contrary to purely mathematical coordinate transformations in SRT. That is why CETs lead to alternative to SRT physics. In particular, CETs predict a dependence of Thomas-Wigner angle $\Omega$ on an "absolute" velocity $\vec{v}$, resulted from the transformation (31).

3) It has been shown that the choice of Galilean transformation in physical space-time within covariant ether theories leads to the Lorentz ether theory. Then kinematics of LET is described by Eqs. (38)-(42). They indicate a principal possibility to measure an absolute velocity experimentally. Therefore, we lose an ether formulation of special relativity principle in the sense adopted by Lorentz and Poincaré. In contrast, we arrive at the conclusion: if an existence of absolute frame is assumed, that an absolute velocity can be detected experimentally, at least in principle.

4) The dependence $\Omega(\vec{v})$ can be measured in the experiments, based on the recent methodological achievements in the Mössbauer spectroscopy. The proposed experiments test the Einstein relativity principle within the scope of validity of general relativity principle, that qualitatively differs these experiments from another, being performed up-to-date in physics of an empty space-time.

**References**

1. C. Moller. The Theory of Relativity (Clarendon Press, Oxford 1972).
2. C. Moller. In: Astrophysics, Quanta and Relativity (Nauka, Moscow, 1987) pp. 17-41 (in Russian).
3. A.A. Logunov. Theory of Relativity and Gravitation (modern analysis of the problem), (Nauka, Moscow, 1987) (in Russian).
4. V.A. Fock. The Theory of Spacetime and Gravitation (Nauka, Moscow, 1955) (in Russian).
5. Y.P. Terletskii. Paradoxes in the Theory of Relativity (Plenum, New York, 1968).
6. 7. A.L. Kholmetskii. Physica Scripta **55** (1997) 18.
7. H. Reichenbach. Relativitatscheorie und Erkenntnis a priori (Berlin, 1920).
8. A.L. Kholmetskii. Physica Scripta **67** (2003) 381.
9. A.L. Kholmetskii. Galilean Electrodynamics **14** (Special Issues 2) (2003) 29.
10. D.C. Champeney, G.R. Isaak, A.M. Khan. Nature **198** (1963) 1186.
11. R.A. Chaldea. Lett. Nuovo Cim. **4** (1963) 821.
12. V.G. Nikolenko, A.B. Popov, G.S. Samosvat. JETF **76** (1979) 393 (in Russian).
13. A.L. Kholmetskii. Hyperfine Interactions **126** (2000) 411.
14. A.L. Kholmetskii, W. Potzel, R. Röhlsberger, et al. Hyperfine Interactions **156/157** (2004) 9.
15. A.L. Kholmetskii, O.V. Missevitch. *Mössbauer concentratometers*. (Minsk, Universitetskoe, 1992) (in Russian)